\documentclass{article}[12pt]
\usepackage{amsmath,amssymb}
\usepackage{graphicx}%
\usepackage{xcolor}
\usepackage{verbatim}
\usepackage{multirow}

\newtheorem{stat}{Statement}

\newcommand{\dt}[1]{\frac{d#1}{dt}}

\newcommand{\TwoFigsReg}[6]{%
\begin{flushleft}
\begin{tabular}{cc}
\parbox{7.5cm}{\centerline{\includegraphics[width=7cm,height=#2cm]{#1}}}  & \parbox{7.5cm}{\centerline{\includegraphics[width=7cm,height=#4cm]{#3}}}  \\
\parbox{7.5cm}{\vspace{7pt}\refstepcounter{figure}Fig. \thefigure.\quad #5\vfill} & \parbox{7.5cm}{\vspace{7pt}\refstepcounter{figure}Fig. \thefigure.\quad #6\vfill}\\
\end{tabular}
\end{flushleft}
\vspace{7pt}
}

\newcounter{strochka}

\newcounter{spisok}
\begin{document}

\begin{center}
{\bf \Large Yu. G. Ignat'ev\footnote{Institute of Physics, Kazan Federal University, Kremlyovskaya str., 35, Kazan, 420008, Russia; email: yurii.ignatev.1947@yandex.ru} }\\[12pt]
{\bf \Large Evolution of spherical perturbations in the cosmological environment of the Higgs scalar field and an ideal scalar charged fluid} \\[12pt]
\end{center}

\abstract{A mathematical model of the evolution of spherical perturbations in a cosmological ideal scalar-charged fluid with scalar Higgs interaction is constructed. A closed mathematical model of linear spherical perturbations in a cosmological medium of a scalar-charged ideal fluid with scalar Higgs interaction is formulated. It is shown that spherical perturbations of the Friedmann metric are possible only in the presence of an isotropic fluid. At singular points of the background cosmological model, perturbations of the metric do not occur and perturbations are described by a vacuum-field model. Exact ones at singular points of the cosmological system are obtained and it is shown that in the case of a stable singular point of the cosmological system, perturbations of the scalar field represent traveling waves, and in the case of an unstable singular point, perturbations represent exponentially growing standing waves. Using numerical modeling, the formation of a stratified halo in the form of growing standing waves is shown.

\textbf{Key words}: scalar charged plasma, cosmological model, scalar Higgs field, gravitational stability, spherical perturbations.
}


\section*{Introduction}
In \cite{YuI_TMF_1}, in connection with the problem of formation of supermassive black holes -- <<SMBH>> in the early Universe (see, for example, \cite{Zhu} -- \cite{Soliton}), a theory of flat perturbations in a cosmological medium of a scalar-charged ideal fluid with Higgs interaction is constructed. At the same time, unlike the previous works of the Author (see, for example, \cite{Ign_GC21_Un}, \cite{Yu_GC_3_22}), this work does not consider a rigorous model of scalar-charged degenerate fermions, constructed on the basis of microscopic dynamics and, as a result, having an extremely complex and cumbersome mathematical structure, but a phenomenological model of a scalar-charged ideal fluid, reflecting the main features of a rigorous dynamic model, but at the same time mathematically much simpler. The use of this model allowed, firstly, to significantly simplify its numerical integration, and, secondly, to conduct an analytical study of a number of its properties, in particular, to analytically study the gravitational instability of the cosmological system of a scalar neutral liquid with a scalar Higgs field and compare the results with the results of an earlier work by Ya. B. Zeldovich, B. A. Malomed and M. Yu. Khlopov \cite{Khlopov}.

In this paper, we apply the mathematical model \cite{YuI_TMF_1} of an ideal scalar charged fluid to the study of the evolution of spherical perturbations in a cosmological medium using special methods developed in the paper \cite{TMF_23_1}. Note that in \cite{TMF_23_1} the evolution of spherical perturbations was considered on the basis of microscopic dynamics and required the construction of an extremely cumbersome mathematical model. It is the complexity and cumbersomeness of this model that made its analytical study difficult and did not allow a full numerical analysis to be carried out. The use of a sufficiently simple and adequate phenomenological model of a cosmological medium allows us to make significant progress in solving the main problem -- constructing a theory of SMBH formation.

\section{Mathematical models of the cosmological system \newline of an ideal fluid with a Higgs scalar field}
\subsection{Self-consistent system of equations}
So, let us consider the phenomenological model of matter based on the classical Higgs field $\Phi$ and an ideal charged liquid, proposed in \cite{YuI_TMF_1}.
The Lagrange function $L_s$ of the scalar Higgs field $\Phi$ is\footnote{Here and below, Latin letters run through the values  $\overline{1,4}$, Greek letters -- $\overline{1,3}$. The Planck system of units \newline $G=c=\hbar=1$ is used throughout.\label{Plank_units}}
%
%
\begin{eqnarray} \label{L_s}
L_s=\frac{1}{16\pi}(g^{ik} \Phi_{,i} \Phi_{,k} -2V(\Phi)),
\end{eqnarray}
where
\begin{eqnarray}
\label{Higgs}
V(\Phi)=-\frac{\alpha}{4} \left(\Phi^{2} -\frac{m^{2}}{\alpha}\right)^{2}
\end{eqnarray}
-- potential energy of a scalar field, $\alpha$ -- self-action constant, $m$ -- mass of quanta.
The components of the energy-momentum tensor (EMT) of scalar fields with respect to the Lagrange function \eqref{L_s} are equal to:
\begin{eqnarray}\label{T_s}
S^i_{k} =\frac{1}{16\pi}\bigl(2\Phi^{,i}\Phi_{,k}- \delta^i_k\Phi_{,j} \Phi^{,j}+2V(\Phi)\delta^i_k \bigr),
\end{eqnarray}
Further, the energy-momentum tensor of an ideal fluid is equal to:
\begin{equation}\label{T_p}
T^i_{k}=(\varepsilon+p)u^i u_k-\delta^i_k p,
\end{equation}
where $u^i$ is the unit time-like vector of the dynamic velocity of the fluid
\begin{equation}\label{u^2=1}
(u,u)=1.
\end{equation}
We write the equation of the scalar field with the source in the form
\begin{equation}\label{Eq_S} \triangle\Phi+V'_\Phi=-8\pi\sigma\Rightarrow \triangle\Phi+\Phi(m^2-\alpha\Phi^2)=-8\pi\sigma, \end{equation}
where the density of the scalar charge $ \sigma$.

Einstein's equations for the system ``scalar field + ideal fluid'' have the form:
\begin{equation}\label{EQ_Einst}
R^i_k-\frac{1}{2}\delta^i_k R=8\pi (T^i_k+S^i_k) + \delta^i_k \Lambda_0,
\end{equation}
where $\Lambda_0$ is the seed value of the cosmological constant, related to its observed value $\Lambda$, obtained by removing the constant terms in the potential energy, by the relation:
\begin{equation}\label{lambda0->Lambda}
\Lambda=\Lambda_0-\frac{1}{4}\frac{m^4}{\alpha}.
\end{equation}

Calculating the covariant divergence of both parts of the Einstein equations \eqref{EQ_Einst}, we obtain the laws of conservation of energy - the momentum of the system:
\begin{equation}\label{2}
\nabla _{k} T^{ik}-\sigma\nabla^{i} \Phi =0.
\end{equation}

Further, in the proposed phenomenological model of an ideal scalar charged liquid, algebraic relationships are defined between three macroscopic scalars $\sigma$, $p$, and $\varepsilon$ \cite{YuI_TMF_1}
\begin{eqnarray}
\sigma=q^2\beta^2\Phi\rho(\varepsilon); & \displaystyle p=\frac{1}{3}\bigl(\varepsilon-q^2\beta^2\Phi^2\rho(\varepsilon)\bigr),\nonumber
\end{eqnarray}
where $q$ is the scalar charge of particles, $\beta$ is a constant determined by the microscopic state, $\rho(\varepsilon)$ is some scalar function. In particular, in the case of a linear relationship $\rho=\varkappa\varepsilon$, which we will consider further, these relationships take the form:
\begin{eqnarray}\label{s,p,e}
\sigma=e^2\Phi\varepsilon; & \displaystyle p=\frac{1}{3}\varepsilon(1-e^2\Phi^2),
\end{eqnarray}
where
\[ e^2\equiv q^2\beta^2\varkappa\geqslant0.\]
-- dimensionless parameter proportional to the square of the scalar charge of particles.

Thus, the mathematical model of an ideal charged liquid $\mathbf{S}$ is determined by an ordered set of four \emph{fundamental parameters}:
\begin{equation}\label{P}
\mathbf{P}=[\alpha,m,e,\Lambda].
\end{equation}
As shown in \cite{YuI_TMF_1}, the above model satisfies the limiting properties of the charged fermion model obtained from the microscopic dynamics of \cite{TMF_23_1}.
In particular, for $e=0$ we obtain from this model a model of the system $\mathbf{S^{(0)}}$ consisting of a scalar Higgs field and a scalar neutral ideal ultrarelativistic liquid
$\sigma=0,p=\varepsilon/3$: $\mathbf{P}=[\alpha,m,0,\Lambda]$ $\equiv \mathbf{P^{(0)}}=[\alpha,m,\Lambda]$. For $\varepsilon\equiv0$ we obtain the \emph{vacuum-field model} $\mathbf{S^{(00)}}$ of a system consisting of a single Higgs field. Note that the algebraic relations \eqref{s,p,e} could be complicated by extending them also to the limiting case of a nonrelativistic fluid $e^2\to\infty$ --
\begin{eqnarray}\label{s,p,e_1}
\sigma=\frac{e^2\Phi}{1+e^2\Phi^2}\ \varepsilon; & \displaystyle p=\frac{1}{3}\frac{1}{1+e^2\Phi^2}\ \varepsilon.
\end{eqnarray}
In this case, as in the fermion model, we obtain $p\to0$ for $e^2\to\infty$; $\sigma\to\varepsilon/\Phi$ and $p\to \varepsilon/3$ for $e^2\to0$; $\sigma\to e^2\Phi$, but in this article we will limit ourselves to a simpler model, noting that for small charges $e^2$ the expressions \eqref{s,p,e_1} coincide with \eqref{s,p,e}.
\subsection{Background state}
As a background, we consider the spatially flat Friedmann metric
\begin{eqnarray}\label{ds_0}
ds_0^2=dt^2-a^2(t)(dx^2+dy^2+dz^2)\equiv 
dt^2-a^2(t)[dr^2+r^2(d\theta^2+\sin^2\theta d\varphi^2)],
\end{eqnarray}
and as a background solution, we consider a homogeneous isotropic distribution of matter in which all hydrodynamic functions and the scalar field depend only on the cosmological time $t$:
\begin{equation}\label{base_state}
\Phi=\Phi(t);\; \varepsilon=\varepsilon(t);\; p=p(t);\; u^i=u^i(t).
\end{equation}
It is easy to see that the substitution
\begin{equation}\label{u_0}
u^i=\delta^i_4
\end{equation}
reduces the equations \eqref{2} to a single \emph{material equation}\footnote{Here and below $\dot{f}=\partial f/\partial t$, $f'=\partial f/\partial r$.}
\begin{equation}\label{7a1-0}
\dt{\varepsilon}+3\frac{\dot{a}}{a}(\varepsilon+p)=\sigma\dot{\Phi};
\end{equation}
In this case, the EMT of the scalar field in the background state also takes the form of the EMT of an ideal isotropic liquid:
\begin{equation} \label{MET_s}
S^{ik} =(\varepsilon_s +p_{s} )u^{i} u^{k} -p_s g^{ik} , \end{equation}
where:
\begin{eqnarray}\label{Es} \varepsilon_s=\frac{1}{8\pi}\biggl(\frac{1}{2}\dot{\Phi}^2+V(\Phi)\biggr);\\ \label{Ps} p_{s}=\frac{1}{8\pi}\biggl(\frac{1}{2} \dot{\Phi}^2-V(\Phi)\biggr), \end{eqnarray}
so:
\begin{equation}\label{e+p}
\varepsilon_s+p_{s}=\frac{1}{8\pi}\dot{\Phi}^2.
\end{equation}
The equations of the background scalar field \eqref{Eq_S} in the Friedmann metric take the form:
\begin{eqnarray}\label{Eq_Phi_eta}
\ddot{\Phi}+\frac{3}{a}\dot{a}\dot{\Phi}+m_0^2\Phi-\alpha\Phi^3= -8\pi\sigma.
\end{eqnarray}
Taking into account the connections \eqref{s,p,e} we obtain the complete system of background equations of the model $\mathbf{S}$ (see \cite{YuI_TMF_1})
\begin{eqnarray}\label{Dxi/dt0}
\dot{\xi}=H;\\
\label{dPhi/dt0}
\dot{\Phi}=Z;\;\\
\label{dH/dt_0}
\dot{H}=- \frac{Z^2}{2}-\frac{4\pi}{3}\varepsilon(4-e^2\Phi^2);\\
\label{dZ/dt0}
\dot{Z}=-3HZ-\Phi(m^2 -\alpha\Phi^2)-8\pi e^2\Phi\varepsilon;\\
\label{de/dt0}
\dot{\varepsilon}=\displaystyle -\varepsilon(4-e^2\Phi)H+e^2\Phi Z\varepsilon,
\end{eqnarray}
where the Hubble parameter $H(t)$ is introduced
\begin{equation}\label{H}
H= \frac{\dot{a}}{a}\equiv \dot{\xi}.
\end{equation}
In this case, Einstein's equation $^4_4$, which is the first integral of the system \eqref{Dxi/dt0} -- \eqref{de/dt0} with a zero value of the constant, takes the form
\begin{eqnarray}\label{Surf_Einst}
3H^2-\frac{Z^2}{2}-8\pi\varepsilon+\frac{\alpha}{4}\left(\Phi^2-\frac{m^2}{\alpha}\right)^2-\Lambda_0=0.
\end{eqnarray}

The background systems of dynamic equations for the $\mathbf{S^(0)}$ and $\mathbf{S^(00)}$ models are obtained from the system \eqref{Dxi/dt0} -- \eqref{de/dt0} by limiting passages $e^2=o0$ and $\varepsilon\equiv0$, respectively.
A qualitative analysis of the background systems and their numerical modeling are carried out in \cite{YuI_TMF_1}.

\section{Linear spherical perturbations of the cosmological model}
\subsection{Spherical perturbations}
We write the metric with gravitational perturbations in isotropic spherical coordinates with a conformally Euclidean metric of three-dimensional
space \footnote {see, for example, \cite{Land_Field}}, which allows a continuous transition to the Friedmann metric \eqref{ds_0} \cite{TMF_23_1}:
\begin{eqnarray}
\label{metric_pert}
ds^2=\mathrm{e}^{\nu(r,t)}dt^2-a^2(t)\mathrm{e}^{\lambda(r,t)}[dr^2+r^2(d\theta^2+\sin^2\theta d\varphi^2)],
\end{eqnarray}
where $\nu(r,t)$ and $\lambda(r,t)$ are small longitudinal perturbations of the Friedman metric ($\nu\ll1,\ \lambda\ll1$).

Assuming further for perturbations of the scalar field, energy density and fluid velocity vector
\begin{eqnarray}\label{delta_Phi}
\Phi(r,t)=\Phi_0(t)+\phi(r,t); & \displaystyle u^i=e^{-\frac{\nu(r,t)}{2}}\delta^i_4+\frac{\upsilon(r,t)}{a^2}\delta^i_1; & \displaystyle
\varepsilon(r,t)=\varepsilon_0(t)+\delta\varepsilon(r,t),
\end{eqnarray}
where $\Phi_0(t)$, $\varepsilon_0(t)$ (as well as $p_0(t)$ and $\sigma_0(t)$) are unperturbed (background) values of the corresponding quantities considered in the previous section, and $\phi(r,t)$, $\upsilon(r,t)$ and $\delta\varepsilon(r,t)$ are their small perturbations. Using formulas \eqref{s,p,e}, we obtain expressions for perturbations of the scalar charge density and pressure\footnote{In what follows, to simplify notation, we will omit the arguments of the functions.}:
\begin{eqnarray}\label{d_sigma}
\sigma(r,t)=\sigma_0(t)+\delta\sigma(r,t) & \sigma_0= e^2\Phi_0\varepsilon_0; & \delta\sigma= e^2(\phi\varepsilon_0+\Phi_0\delta\varepsilon);\nonumber\\
\label{d_p}
p(r,t)=p_0(t)+\delta p(r,t); & \displaystyle p_0=\frac{1}{3}\varepsilon_0(1-e^2\Phi^2_0); & \displaystyle \delta p=\frac{1}{3}[\ \delta\varepsilon(1-e^2\Phi^2_0)-2e^2\varepsilon_0\Phi_0\phi\ ];\nonumber\\
\delta\varepsilon+\delta p= & \displaystyle \frac{1}{3}[\ \delta\varepsilon(4-e^2\Phi^2_0)-2e^2\varepsilon_0\Phi_0\phi\ ]. &
\end{eqnarray}
\subsection{Equations for model perturbations}
\subsubsection{Perturbations of the total energy-momentum tensor}
Thus, non-zero perturbations of the components of the total energy-momentum tensor have the form:
\begin{eqnarray}\label{TS^a_a}
8\pi\delta(S^\beta_\beta+T^\beta_\beta)=\phi\Phi_0(m^2-\alpha\Phi_0^2)-\dot{\phi}\dot{\Phi_0}+\frac{1}{2}\nu\dot{\Phi}_0^2-8\pi \delta p, & (\beta=\overline{1,3});\\ \label{FS^4_4} 8\pi\delta(S^4_4+T^4_4)=\phi\Phi_0(m^2-\alpha\Phi_0^2)+\dot{\phi}\dot{\Phi_0}-\frac{1}{2}\nu\dot{\Phi}_0^2-8\pi \delta\varepsilon; &\\
\label{TS^1_4}
8\pi\delta(S^1_4+T^1_4)=\dot{\Phi}_0\phi'-\upsilon(\varepsilon_0+p_0)\Rightarrow & \displaystyle -\frac{1}{a^2}\delta(S^1_4+T^1_4)=\delta(S^4_1+T^4_1),
\end{eqnarray}
where it is necessary to substitute the expression for the pressure perturbation $\delta p$ from \eqref{d_p}.

\subsubsection{Perturbation of the field equation}
\begin{eqnarray}\label{dEq_phi}
\ddot{\phi}+3H\dot{\phi}-\frac{1}{a^2r^2}\frac{\partial}{\partial r}r^2\frac{\partial\phi}{\partial r}+\phi(m^2-3\alpha\Phi_0^2)+\frac{1}{2}\dot{\Phi}_0(3\dot{\lambda}-\dot{\nu})
+\nu(m^2\Phi_0-\alpha\Phi_0^3+8\pi\sigma_0)=-8\pi\delta\sigma,
\end{eqnarray}
where the expression for the pressure perturbation $\delta\sigma$ must be substituted from from \eqref{d_sigma}.

\subsubsection{Perturbed Einstein equations\label{flat_pert_Einst}}
Non-trivial perturbed Einstein equations have the form:
\begin{eqnarray}\label{d11-d22}
\delta(EQ^1_1-EQ^2_2): & \displaystyle \frac{r}{2a^2}\frac{\partial}{\partial r}\left(\frac{1}{r}\frac{\partial}{\partial r}(\lambda+\nu)\right)=0\Rightarrow \frac{\partial}{\partial r}(\lambda+\nu)= C_1r\Rightarrow
\lambda+\nu=\frac{1}{2}C_1r^2+C_2;\\
\label{d11}
\delta EQ^1_1: & \displaystyle \ddot{\nu}+4H\dot{\nu}+\left(3H^2-\frac{Z^2}{2}+8\pi(\varepsilon_0+p_0)\right)\nu+Z\dot{\phi}-\Phi_0(m^2-\alpha\Phi_0^2)\phi=-8\pi\delta p;\\
\label{d44}
\delta EQ^4_4: & \displaystyle \frac{1}{a^2r^2}\frac{\partial}{\partial r}\left(r^2\frac{\partial\nu}{\partial r}\right) -3H\dot{\nu}-Z\dot{\phi}-\left(3H^2-\frac{1}{2}Z^2\right)\nu-\Phi_0(m^2-\alpha\Phi_0^2)\phi=8\pi\delta\varepsilon;\\
\label{d14}
\delta EQ^4_1: & \displaystyle 8\pi (\varepsilon_0+p_0)\upsilon-(\dot{\Phi}_0\phi+H\nu-\dot{\lambda})'=0.
\end{eqnarray}
\subsubsection{General properties of the system}
First, note that due to the conditions for the transition of the system's solutions to the background solution at spatial infinity, the zero boundary conditions must be satisfied
\begin{eqnarray}
\left.\lambda(r,t)\right|_{r\to\infty}=\left.\nu(r,t)\right|_{r\to\infty}0=\left.\phi(r,t)\right|_{r\to\infty} =0;\\
\left.\lambda'(r,t)\right|_{r\to\infty}=\left.\nu'(r,t)\right|_{r\to\infty}0=\left.\phi'(r,t)\right|_{r\to\infty} =0.
\end{eqnarray}
It follows that in the solution \eqref{d11-d22} $C_1=C_2=0$ should be. Thus:
\begin{equation}\label{nu+lambda=0}
\lambda(r,t)=-\nu(r,t).
\end{equation}
Taking this solution into account, we reduce the field equation \eqref{dEq_phi} and the remaining Einstein equations \eqref{d11} -- \eqref{d14} to the form
\begin{eqnarray}\label{dEq_phi_0}
\ddot{\phi}+3H\dot{\phi}-\frac{1}{a^2r^2}\frac{\partial}{\partial r}r^2\frac{\partial\phi}{\partial r}+\phi(m^2-3\alpha\Phi_0^2)-2Z\dot{\nu}
+\nu(m^2\Phi_0-\alpha\Phi_0^3+8\pi\sigma_0)=-8\pi e^2(\phi\varepsilon_0+\Phi_0\delta\varepsilon);
\end{eqnarray}
\begin{eqnarray}
\label{d11_0}
\delta EQ^1_1: & \displaystyle \ddot{\nu}+4H\dot{\nu}+\left(3H^2-\frac{Z^2}{2}+8\pi(\varepsilon_0+p_0)\right)\nu+Z\dot{\phi}-\Phi_0(m^2-\alpha\Phi_0^2)\phi \nonumber\\
 & \displaystyle =-\frac{8\pi}{3}[\ \delta\varepsilon(4-e^2\Phi^2_0)-2e^2\varepsilon_0\Phi_0\phi\ ];\\
\label{d44_0}
\delta EQ^4_4: & \displaystyle \frac{1}{a^2r^2}\frac{\partial}{\partial r}\left(r^2\frac{\partial\nu}{\partial r}\right) -3H\dot{\nu}-Z\dot{\phi}-\left(3H^2-\frac{1}{2}Z^2\right)\nu-\Phi_0(m^2-\alpha\Phi_0^2)\phi=8\pi\delta\varepsilon;\\
\label{d14_0}
\delta EQ^4_1: & \displaystyle 8\pi (\varepsilon_0+p_0)\upsilon-(Z\phi+H\nu+\dot{\nu})'=0.
\end{eqnarray}

Thus, we have four linear differential equations in four unknown functions $\nu(r,t)$, $\phi(r,t)$, $\delta\varepsilon(r,t)$ and $\upsilon(r,t)$, i.e., a completely defined system of equations.
Secondly, we note that the equation \eqref{d14} for $\varepsilon_0+p_0\not\equiv0$ is in fact the definition of the magnitude of the velocity of matter $\upsilon(\eta)$. Otherwise, if there is no ideal fluid, i.e., \emph{in the vacuum-field model} $\mathbf{S^{(00)}}$, from \eqref{d14_0} we obtain
\begin{equation}\label{e+p=0}
\varepsilon_0+p_0\equiv0 \Rightarrow Z\phi+H\nu+\dot{\nu} =f(t),
\end{equation}
where $f(t)$ is an arbitrary function. Using this relationship in the remaining equations for the perturbations \eqref{dEq_phi_0} -- \eqref{d44_0}, we obtain 3 independent linear differential equations with respect to one unknown function. Thus, the system is essentially overdetermined and, generally speaking, inconsistent.

Thus, the following statement is true, similar to the statement in the article \cite{YuI_TMF_1}.
\begin{stat}\label{epsilon_not0}\hskip -4pt \textbf{.}

Spherical perturbations of the Friedmann metric \eqref{metric_pert} -- \eqref{delta_Phi} in the model with a scalar field are possible only in the presence of an isotropic fluid.

\end{stat}

Thus, assuming $\varepsilon_0\not\equiv0$, we obtain from \eqref{d14_0} the definition of the radial velocity:
\begin{eqnarray}\label{v=}
\upsilon=\frac{\partial}{\partial r}\frac{Z\phi+H\nu+\dot{\nu}}{8\pi (\varepsilon_0+p_0)}.
\end{eqnarray}
Thus, we are left with 3 second-order linear differential equations \eqref{dEq_phi_0} -- \eqref{d44_0} in three functions $\nu(r,t)$, $\phi(r,t)$ and $\delta\varepsilon(r,t)$, i.e., we have a
completely defined system.

\section{Asymptotic properties of solutions at singular points of a cosmological system}
\subsection{Equations for perturbations in the neighborhood of singular points}
In \cite{YuI_TMF_1} it is shown that the singular points $M(H_0,\Phi_0,Z_0,\varepsilon_0)$ of the dynamical system \eqref{dPhi/dt0} -- \eqref{de/dt0}, describing the unperturbed state, are:
\begin{eqnarray}\label{M_0}
M^\pm_0=\left[\pm \sqrt{\frac{\Lambda}{3}},0,0,0\right];&\displaystyle M^\pm_\pm=\left[\pm \sqrt{\frac{\Lambda_0}{3}},\pm\frac{m}{\sqrt{\alpha}},0,0\right],
\end{eqnarray}
where the cosmological constants $\Lambda$ and $\Lambda_0$ are related by the relation \eqref{lambda0->Lambda}, and always $\Lambda<\Lambda_0$. In this case, only the singular point $M^+_0$ is stable (attractive center), the other points are unstable -- saddle or repulsive.

Assuming for now that the general conditions are satisfied at the singular points
\begin{eqnarray}\label{sing_points}
\Phi_0(m^2-\Phi_0^2)=0; Z=0; &\displaystyle H=H_0=\mathrm{Const}; & \varepsilon_0=0,
\end{eqnarray}
note that at these points, due to \eqref{d_sigma}, also
\[p_0=\sigma_0=\delta p=0.\]
However, $\delta\varepsilon,\delta\sigma$, generally speaking, are not equal to zero --
\[\delta\sigma=e^2\Phi_0\delta\varepsilon\]
-- $\delta\sigma$ vanishes at a stable point of the system, and is nonzero at an unstable point.

Further, if the condition \eqref{sing_points} is satisfied, the unperturbed equations of the gravitational and scalar fields are reduced to the only nontrivial Einstein equation \eqref{Surf_Einst}
\begin{equation}\label{Eq_a(t)}
\Phi_0=0 \Rightarrow \frac{\dot{a}^2}{a^2}= \displaystyle\frac{\Lambda}{3}; \qquad
\displaystyle \Phi_0=\pm \frac{m}{\sqrt{\alpha}} \Rightarrow \frac{\dot{a}^2}{a^2}=\frac{\Lambda_0}{3}.
\end{equation}
Thus, we obtain the well-known inflationary solution for the scale factor (see, for example, \cite{Weinberg})
\begin{eqnarray}\label{a(t)}
a(t)=\mathrm{e}^{\pm H_0t}: & \displaystyle \Phi_0=0\Rightarrow H_0= \sqrt{\frac{\Lambda}{3}};  & \displaystyle\Phi_0=\pm \frac{m}{\sqrt{\alpha}}\Rightarrow H_0= \sqrt{\frac{\Lambda_0}{3}}.
\end{eqnarray}

Next, from \eqref{e+p=0} taking into account \eqref{sing_points} we obtain $\nu=\nu(t)\Rightarrow \lambda=\lambda(t)$. Using an admissible transformation of the time variable and the scale factor, we achieve $\nu(t)=0$. Therefore, in this case, there are no metric perturbations. But then the equations \eqref{d11_0} and \eqref{d44_0} give $\delta\varepsilon=0$. Thus, all the equations of the system \eqref{d11_0} -- \eqref{d14_0} turn into identities and only one equation remains with respect to the perturbation of the scalar field \eqref{dEq_phi_0}, which takes the form:
\begin{eqnarray}\label{dEq_phi_00}
\ddot{\phi}+3H\dot{\phi}-\frac{1}{a^2r^2}\frac{\partial}{\partial r}r^2\frac{\partial\phi}{\partial r}+\phi(m^2-3\alpha\Phi_0^2)=0.
\end{eqnarray}

Thus, the following statement is true.
\begin{stat}\label{Vac_Field}\hskip -4pt \textbf{.}

At singular points of the background cosmological model \eqref{sing_points}, gravitational perturbations do not arise, and \emph{linear} perturbations of the total EMT \eqref{TS^a_a} -- \eqref{TS^1_4} vanish regardless of the magnitude of the perturbation of the scalar potential $\phi(\eta)$, relative to which only one field equation \eqref{dEq_phi_00} remains. Thus, in the vicinity of singular points of the background, perturbations are described by the vacuum-field model
$\mathbf{S^{(00)}}$.

\end{stat}
\subsection{Exact solution}
Substituting the solution \eqref{a(t)} into the equation \eqref{dEq_phi_00} and separating the variables ($\phi(r,t)=R(r)T(t)$), we obtain the equations :
\begin{eqnarray}\label{eq_T}
\ddot{T}\pm 3H_0\dot{T}+\left(m^2-3\alpha\Phi_0^2+n^2e^{\mp2H_0t}\right)T=0;\\
\label{eq_R} R''+\frac{2}{r}R'+n^2R=0,
\end{eqnarray}
where $(-n^2)\leqslant 0$ is the separation constant. By solving these equations, we find their general solutions;
\begin{eqnarray}\label{RT=}
R(r)=C_1\frac{\sin(nr)}{r}+C_2\frac{\cos(nr)}{r};& \displaystyle T(t)=e^{-3/2H_0t}\left[\tilde{C}_1 \mathrm{J}_\gamma\left(n\frac{\mathrm{e}^{-H_0t}}{H_0}\right)
+\tilde{C}_2 \mathrm{Y}_\gamma\left(k\frac{\mathrm{e}^{-H_0t}}{H_0}\right)\right],
\end{eqnarray}
where $\mathrm{J}_\gamma(z)$ and $\mathrm{Y}_\gamma(z)$ -- Bessel functions of the first and second kind with index $\gamma$ (see, for example, \cite{Lebedev}):
\[\gamma\equiv -\frac{3}{2}\left(1-\frac{4}{9H^2_0}(m^2-3\alpha\Phi_0^2)\right)^{1/2}.\]
According to \eqref{a(t)} for the index $\gamma$ of Bessel functions we obtain
\begin{eqnarray}\label{gamma}
\gamma=\displaystyle -\frac{3}{2}\sqrt{1-\frac{4m^2}{3\Lambda}},& \displaystyle (\Phi_0=0); \qquad \displaystyle \gamma=-\frac{3}{2}\sqrt{1+\frac{8m^2}{3\Lambda_0}}, &
\displaystyle \bigl(\Phi_0=\pm\frac{m}{\sqrt{\alpha}}\bigr) .
\end{eqnarray}

First, note that by moving to a conformal time variable $\eta$
\[ad\eta=dt\Rightarrow \eta=-\frac{1}{H_0}\mathrm{e}^{-H_0t}\]
and extending the values of $k$ to the entire real axis $k\in(-\infty,+\infty)$, we can write the particular solutions $\phi(r,t)$ in a more compact form:
\begin{eqnarray}\label{phi_n=}
\phi_n(r,\eta)=\frac{\mathrm{e}^{inr}}{r}\ (-\eta)^{3/2}[\mathrm{C}^1_n\mathrm{J}_\gamma(n\eta)+\mathrm{C}^2_n\mathrm{Y}_\gamma(n\eta)] .
\end{eqnarray}
When replacing $z\leftrightarrow r$, this solution formally differs from the corresponding plane-wave \cite{YuI_TMF_1} only by the factor $1/r$. Then the general solution is determined by the Fourier integral
\begin{eqnarray}\label{phi(r,t)=}
\phi(r,t)=\frac{(-\eta)^{3/2}}{2\pi r}\int\limits_{-\infty}^{+\infty}\mathrm{e}^{inr}[\mathrm{C}^1_n\mathrm{J}_\gamma(n\eta)+\mathrm{C}^2_n\mathrm{Y}_\gamma(n\eta)]\ dn.
\end{eqnarray}
First, note that the singular part of the solution \eqref{phi(r,t)=} is proportional to the central scalar charge $Q$, the constant $C_2$ in the particular solution \eqref{RT=} is equal to $Q(n)$.
Finally, third, note that in the case $\Phi_0=0$, when the condition $3\Lambda-4m^2<0$ is satisfied, the state of the unperturbed vacuum-field model is stable \cite{YuI_TMF_1}, unlike any other cases. Therefore, it is in the case of a stable vacuum-field model that the index $\gamma$ of the Bessel functions in the solutions \eqref{phi_n=} and \eqref{phi(r,t)=} is purely imaginary $\gamma^2<0$ \cite{YuI_TMF_1}.

\subsection{Numerical simulation of the vacuum-field model}
Below we consider two examples of solutions \eqref{phi_n=} for the values of the problem parameters
\[P = [[\alpha,m,\Lambda_0],[C_1,C_2,n],s],\]
where $s=0$ for $\Phi_0=0$ (stable inflation) and $s=1$ for $\Phi_0=\pm m/\sqrt{\alpha}$ (unstable inflation), the formulas \eqref{a(t)} and \eqref{gamma}:
\begin{eqnarray}\label{P_0,P_1}
P_0 = [[1,1,1],[1,1,1],0];\; P_1 = [1,1,1],[1,1,1],1].
\end{eqnarray}
In this case, the system with parameters $P_0$ is stable, the system with parameters $P_1$ is unstable, and
\[\Lambda=\frac{3}{4};\; \gamma_0=-\frac{i}{2}\sqrt{39};\; \gamma_1=-\frac{9}{2}.\]

\TwoFigsReg{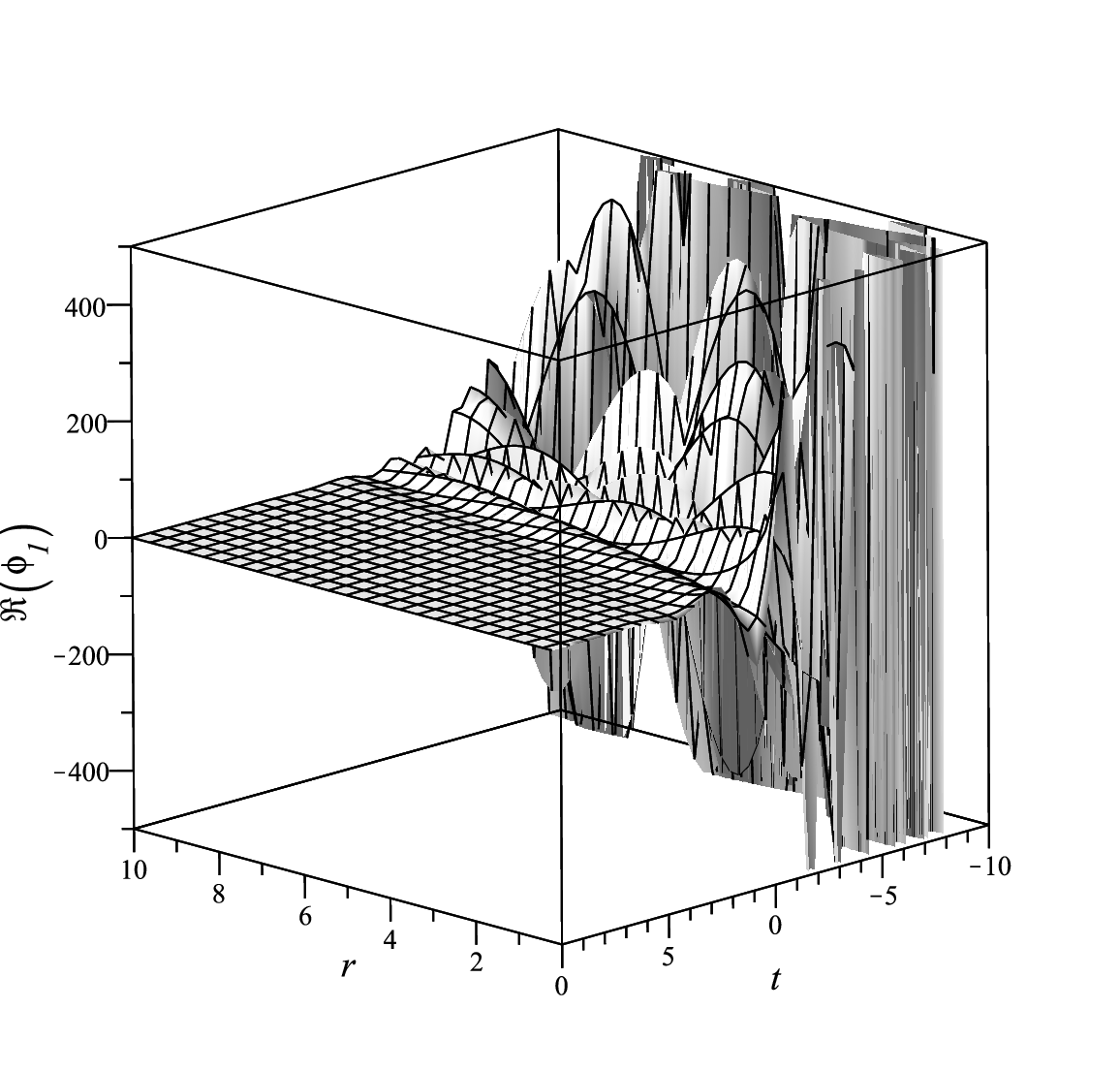}{7}{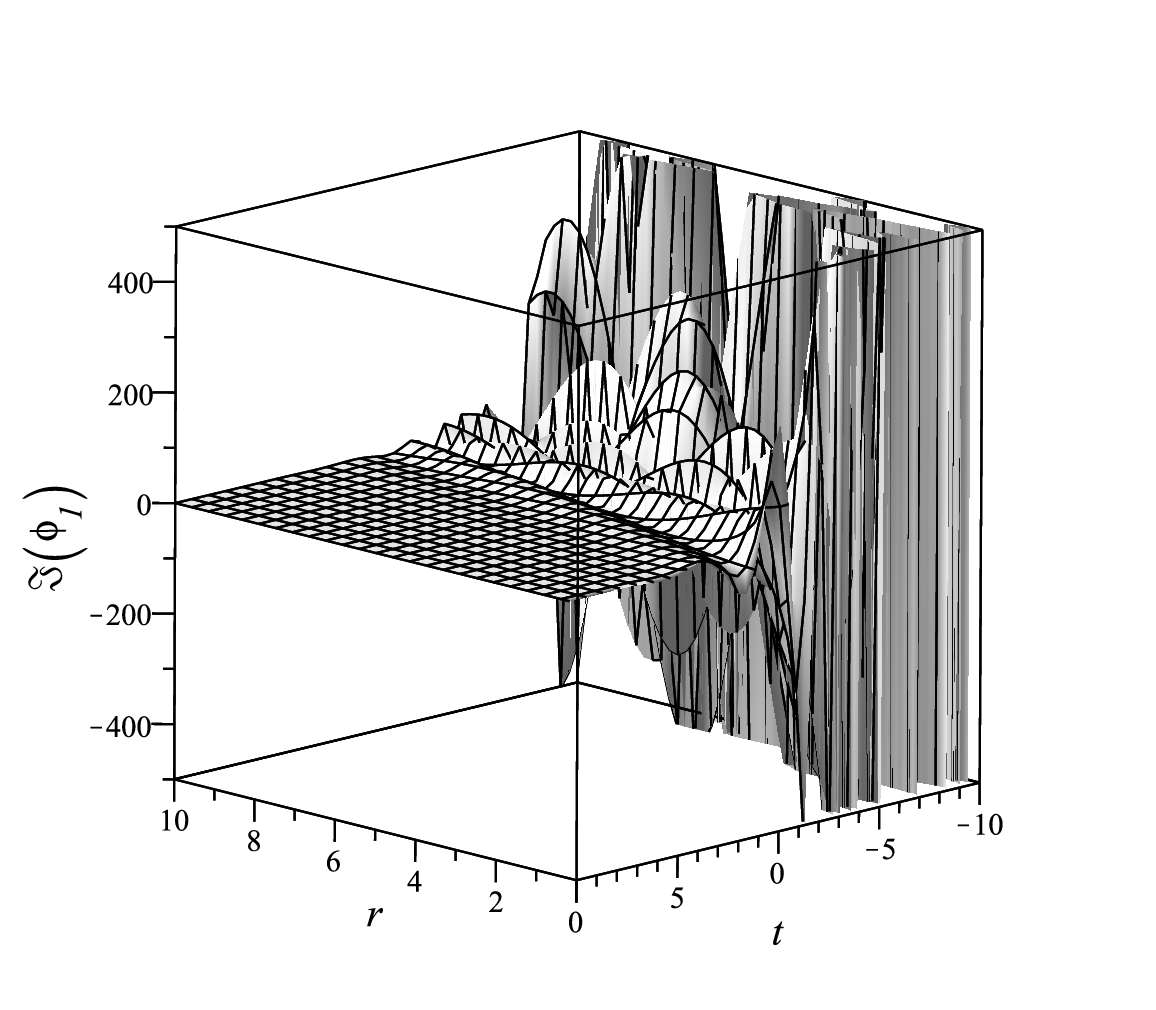}{7}{\label{fig1}Graph of $\Re(\phi_1(r,t))$ in the model $P_0$ \eqref{P_0,P_1} near the center.}{\label{fig2}Graph of $\Im(\phi_1(r,t))$ in the model $P_0$ \eqref{P_0,P_1} near the center.}

Fig. \ref{fig1} and \ref{fig2} show the graphs $\phi(\eta)$ of the real and imaginary parts of the solutions \eqref{phi_n=} for the parameters $P_0$, and Fig. \ref{fig3} and \ref{fig4} show the graphs $\phi(\eta)$ of the real and imaginary parts of the solutions \eqref{phi_n=} for the parameters $P_1$. From these graphs, it follows, firstly, that the real and imaginary parts of the solutions \eqref{phi_n=} behave approximately the same. Secondly, it follows from these graphs that in the case of a stable background solution with infinite amplitude, oscillations in the past disappear in the future, while in the case of an unstable background solution, and in the case of an unstable background solution, $\phi(r,t\to \infty)\to \infty$, i.e., in the case of an unstable background solution, spherical perturbations of the scalar field also become unstable.

\TwoFigsReg{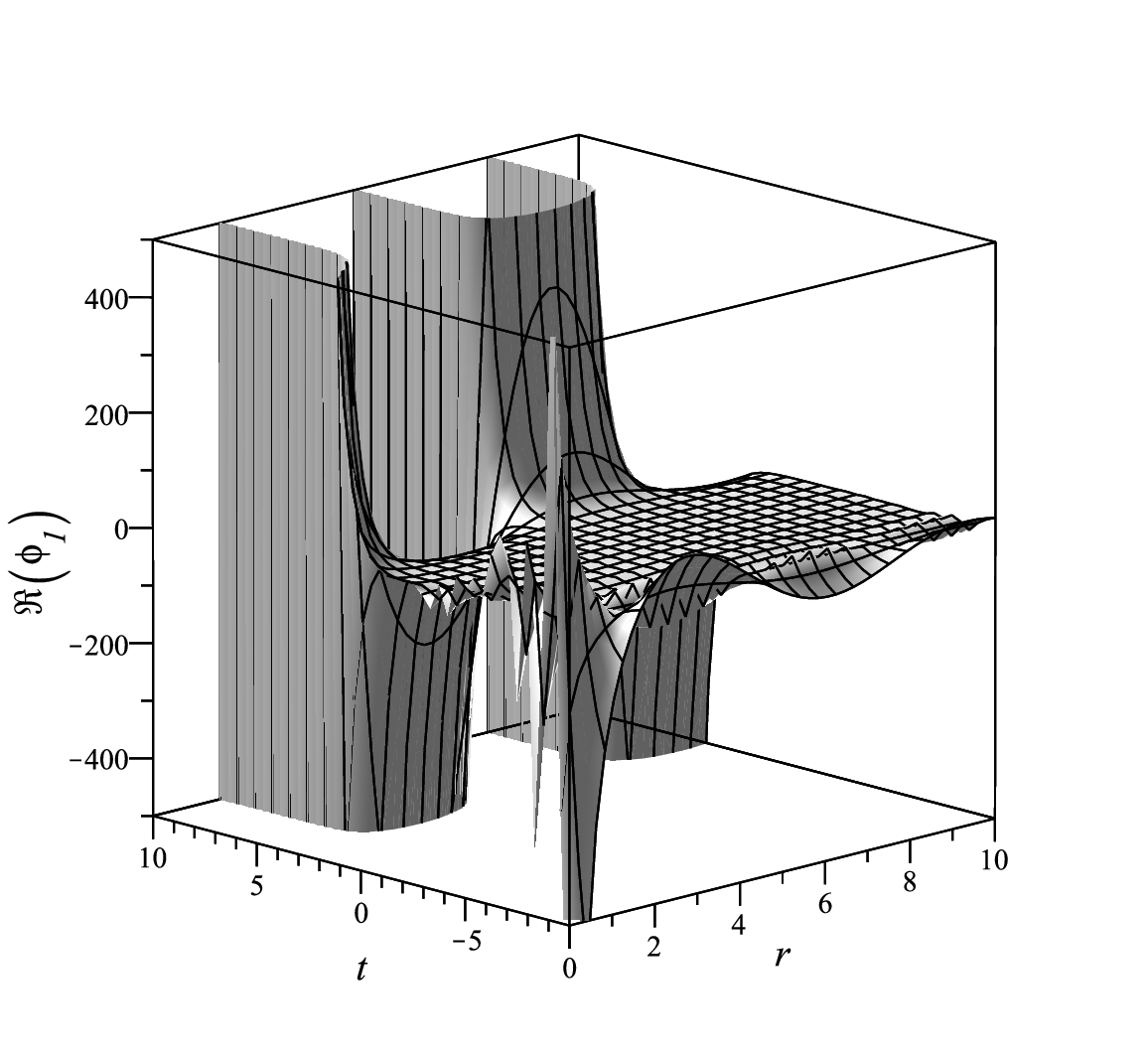}{7}{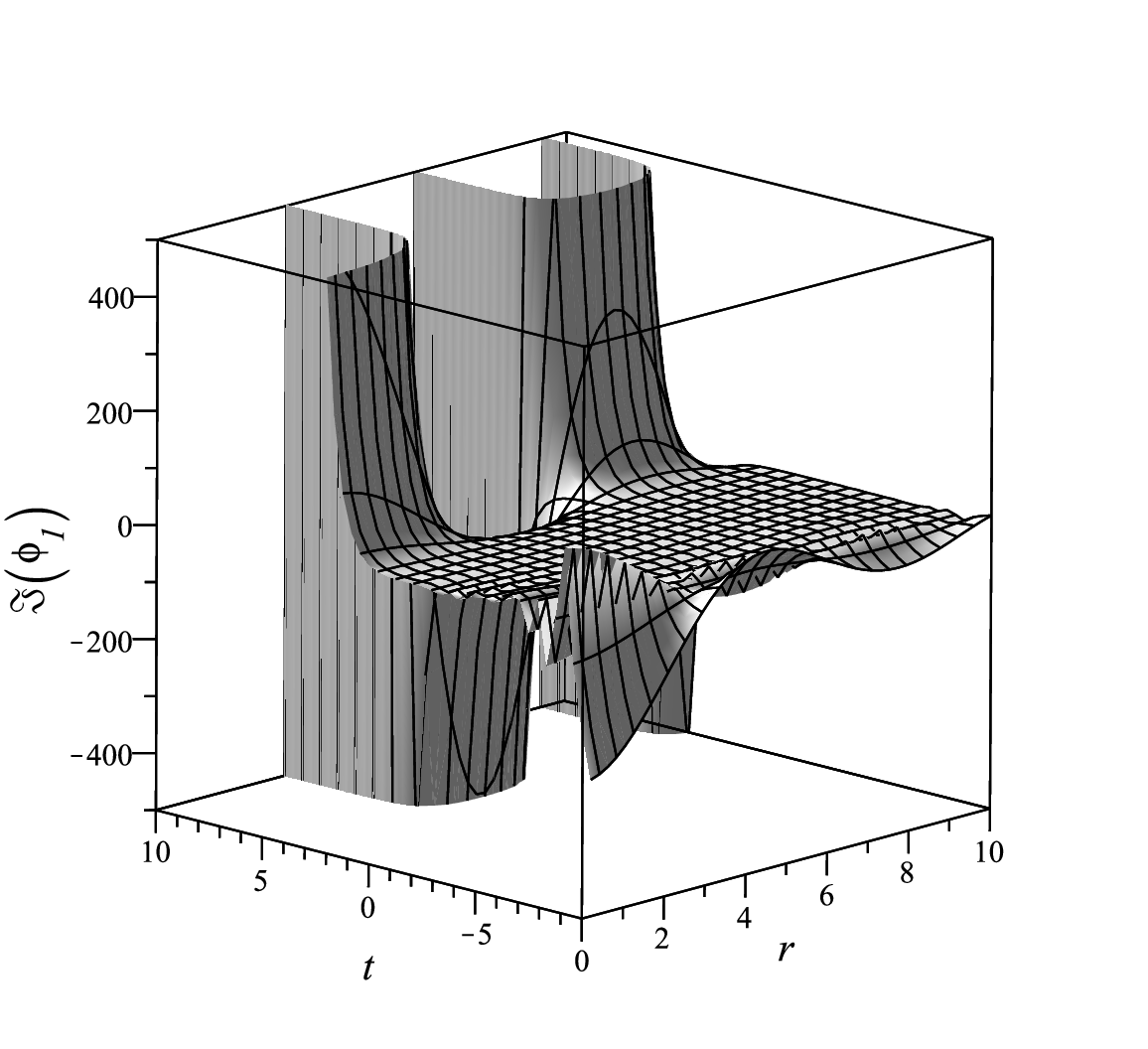}{7}{\label{fig3}Graph of $\Re(\phi_1(r,t))$ in the model $P_1$ \eqref{P_0,P_1} near the center.}{\label{fig4}Graph of $\Im(\phi_1(r,t))$ in the model $P_1$ \eqref{P_0,P_1} near the center.}

Fig. \ref{fig5} -- \ref{fig6} shows the evolution of the pattern of growth of scalar field perturbations in the case of unstable inflation: Fig. \ref{fig5} on the time interval $t\in[0,5]$, Fig. \ref{fig6} -- on the time interval $t\in[5,10]$. Despite the similarity of the graphs, we draw attention to their range of vertical axes: $[-400,+400]$ -- in the first case and $[-2\cdot10^6,+2\cdot10^6]$. Thus, the amplitude of perturbations during this time increases by 4 orders of magnitude. In addition, we draw attention to the steadily exponentially growing spatial structure of perturbations. Note that this structure arises along the radial coordinate $r$. This means that periodic concentric layers with high positive and negative values of the scalar potential are formed around the center.

\TwoFigsReg{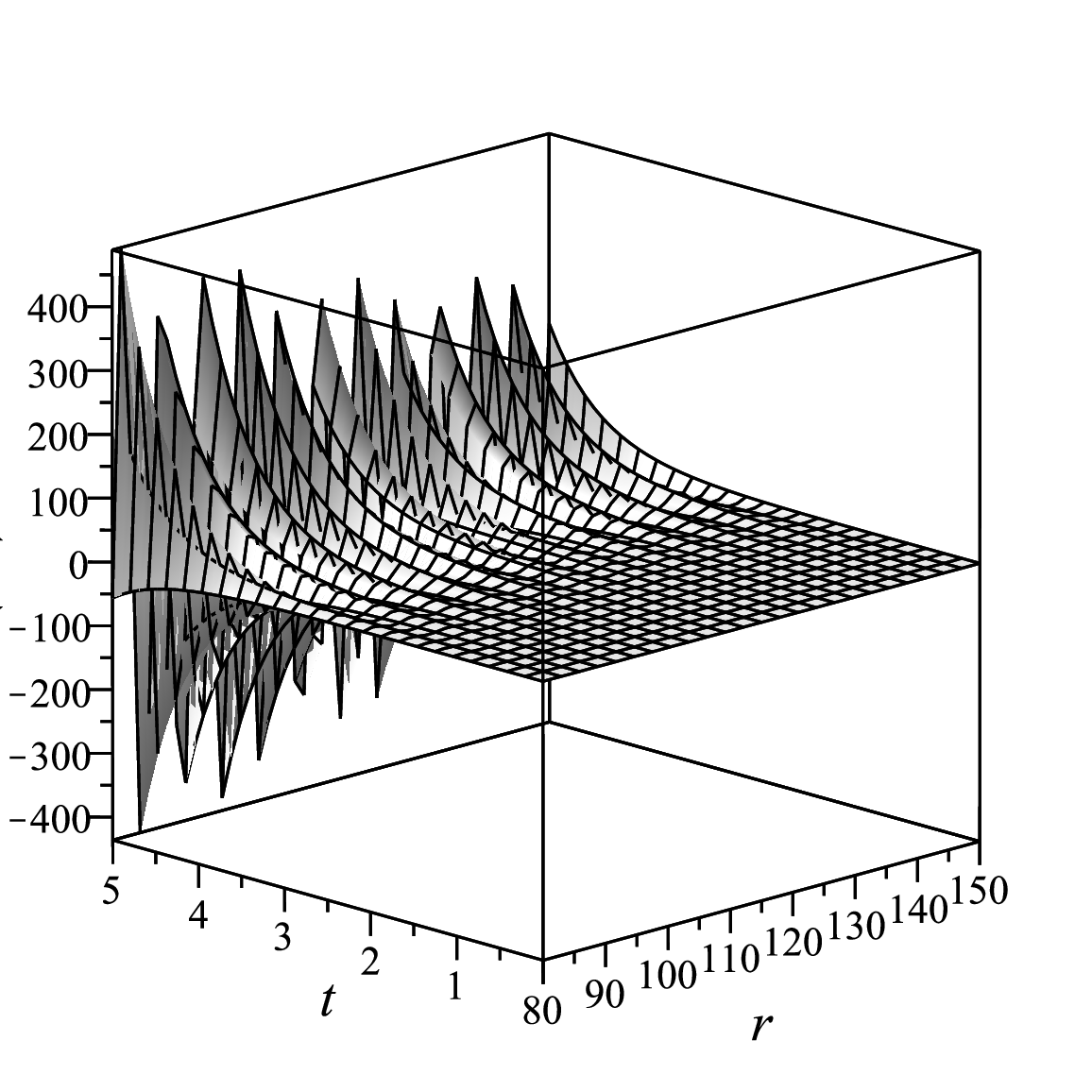}{7}{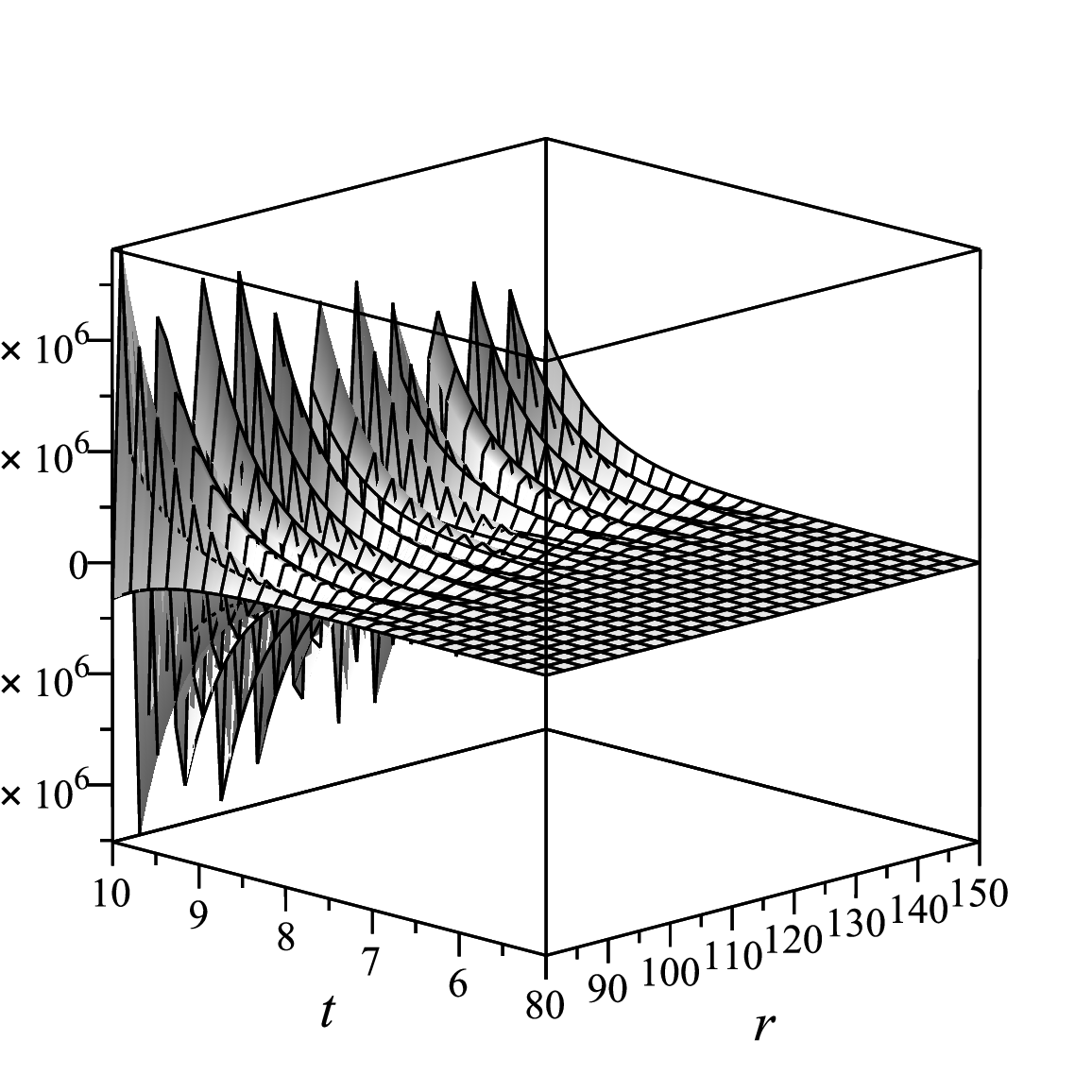}{7}{\label{fig5}Plot of $\Re(\phi_1(r,t))$ in model $P_1$ \eqref{P_0,P_1} away from center $t=\overline{0,5}$.}{\label{fig6}Plot of $\Im(\phi_1(r,t))$ in model $P_1$ \eqref{P_0,P_1} away from center $t=\overline{5,10}$.}

Further, Fig. \ref{fig7} -- \ref{fig8} show graphs of the function $\Re(\phi_1(r,t_0))$ as a function of the radial coordinate for different moments of time (the amplitude of the oscillations increases from smaller to larger moments). The graphs of Fig. \ref{fig7} correspond to stable inflation ($s=0$), and the graphs of Fig. \ref{fig8} -- to unstable inflation ($s=1$). As can be seen, in the case of stable inflation, the oscillations are running waves, while in the case of unstable inflation, they are standing oscillations. In addition, from the graphs in these figures, one can clearly observe a drop in the amplitude of the oscillations with an increase in the radius $\sim 1/r$.

\TwoFigsReg{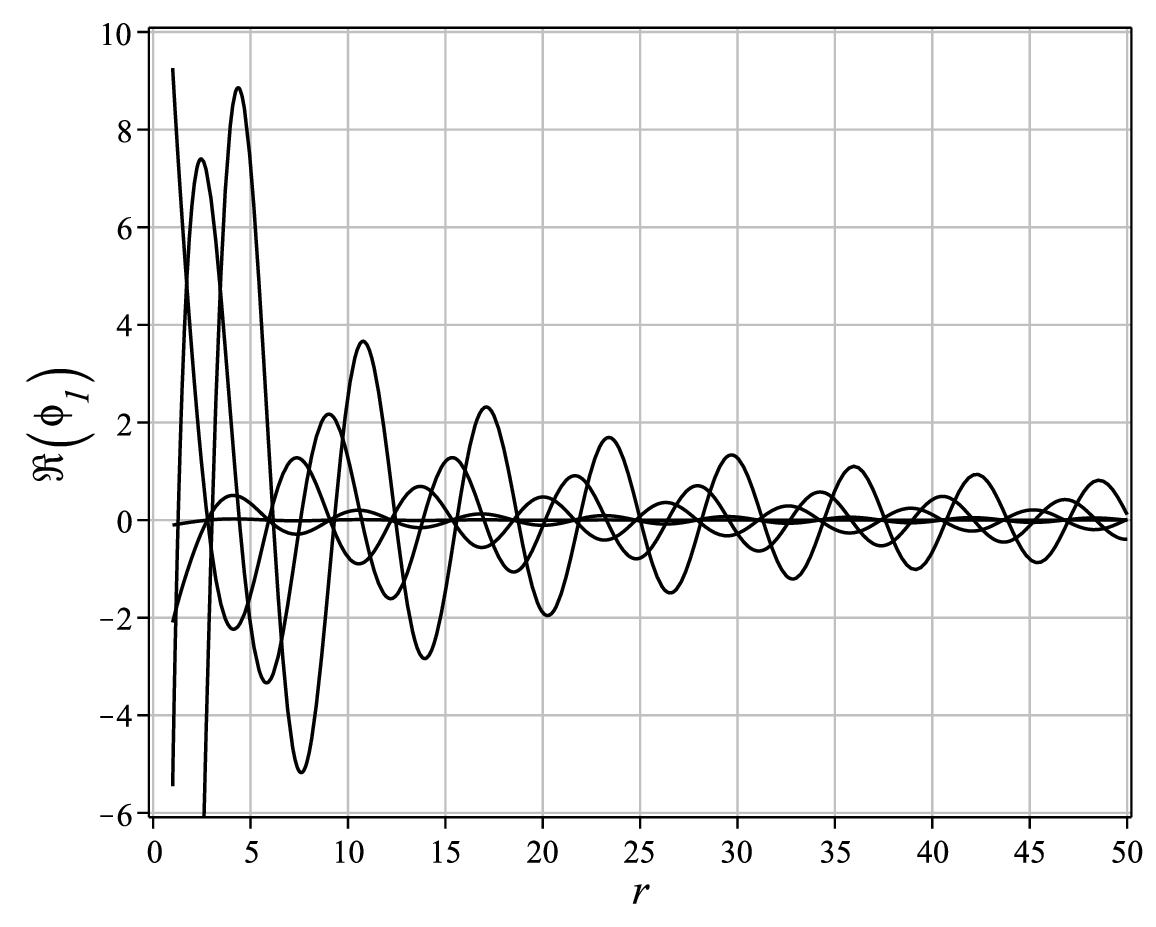}{7}{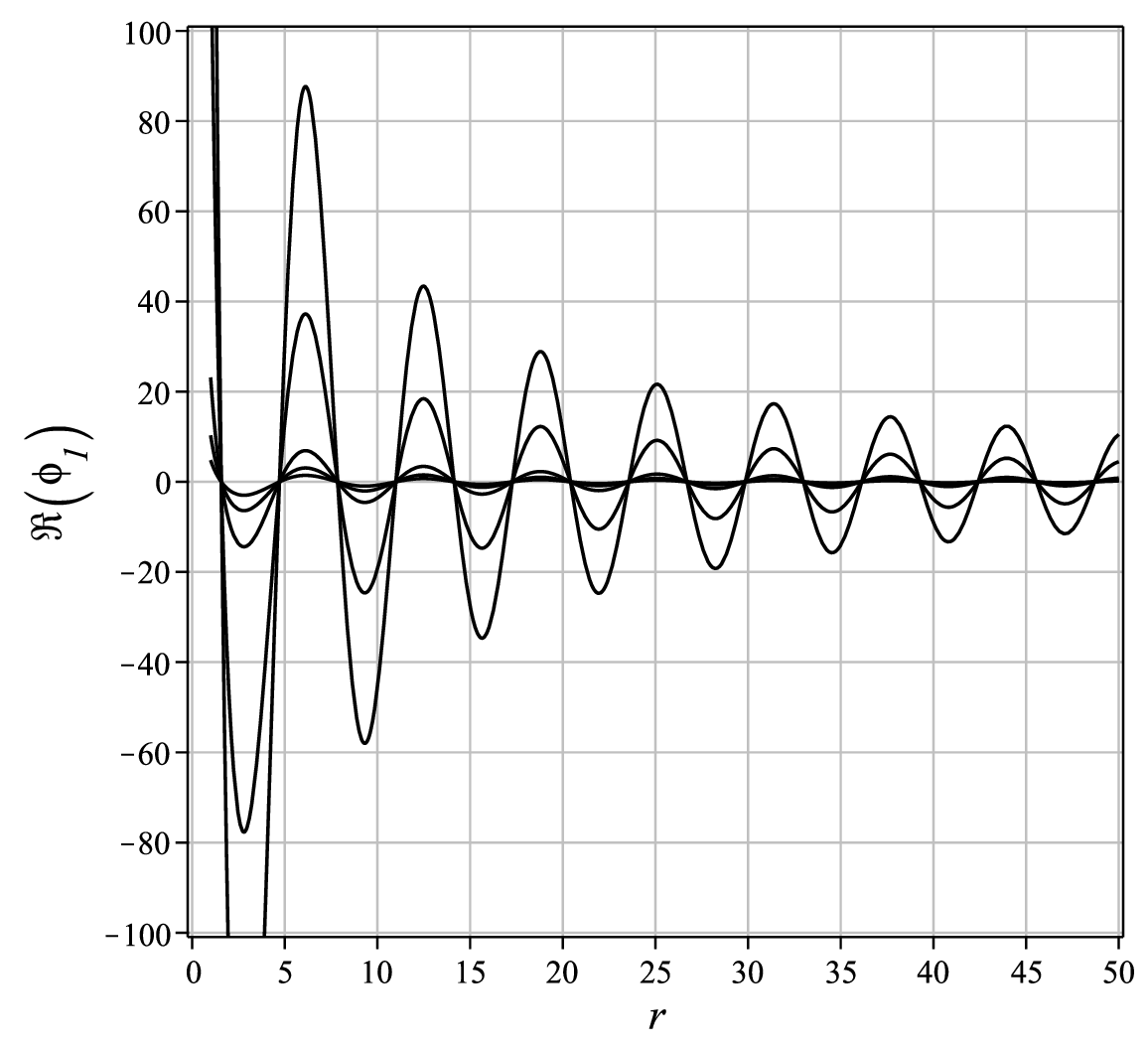}{7}{\label{fig7}Graph of $\Re(\phi_1(r,t_0))$ in model $P_0$ \eqref{P_0,P_1}: $t_0=0;1;2;4;8$.}{\label{fig8}Graph of $\Re(\phi_1(r,t_0))$ in model $P_1$ \eqref{P_0,P_1}: $t_0=0;0.5;1;2;82.5$.\\}
\section*{Conclusion}
To summarize the work, we highlight its most significant results.

\begin{itemize}
\item A closed mathematical model of linear spherical perturbations in the cosmological medium of a scalar-charged ideal fluid with scalar Higgs interaction is formulated.
\item It is shown that spherical perturbations of the Friedmann metric \eqref{metric_pert} -- \eqref{delta_Phi} in the model with a scalar field are possible only in the presence of an isotropic fluid.
\item It is shown that at singular points of the background cosmological model, perturbations of the metric do not arise, i.e., in the vicinity of singular points of the background, perturbations are described by the vacuum-field model.
\item Exact solutions of the vacuum-field model at singular points of the cosmological system are obtained and it is shown that in the case of a stable singular point of the cosmological system, the perturbations of the scalar field represent traveling waves,
and in the case of an unstable singular point, the perturbations represent exponentially growing standing waves.
\end{itemize}

Thus, it is in the case of stable inflation that growing spherical standing waves are formed. Note that in the work \cite{YuI_TMF_1} a similar property was noted with respect to plane perturbations of the cosmological system of an ideal scalar charged fluid. Also note that the possibility of the formation of growing standing waves was first indicated in the work of M. Yu. Khlopov, B. A. Melomed and Ya. B. Zeldovich \cite{Khlopov}.

For convenience of presentation of the results, Fig. \ref{fig9} -- \ref{fig12} show these oscillation modes in spherical coordinates in the gradient graphics format, wherein Fig. \ref{fig9} and \ref{fig11} show these modes for the case of stable inflation $P_0$ (the late stage of the cosmological expansion), and Fig. \ref{fig9} and \ref{fig11} show these modes for the case of unstable inflation $P_1$ (the early stage of the cosmological expansion). Commenting on these graphs, we note that their light areas correspond to the regions $r$ with the maximum positive value of perturbations of the potential $\Re(\phi_1(r,t))$, and the darkest areas correspond to the regions $r$ with the maximum negative value of perturbations of the potential $\Re{\phi_1(r,t)}$, so that brighter and more contrasting images indicate a larger amplitude of the standing waves. Thus, the graphs in Fig. \ref{fig9} -- \ref{fig12} demonstrate
the formation of a spherical halo, stratified by the structure of standing waves growing with time (until the end of early inflation).

In this regard, we note the works \cite{Supermass_BH}, \cite{Shadows}, \cite{Soliton}, which consider the possibility of the existence of \emph{scalar halos} and \emph{scalar hair} in the vicinity of supermassive black holes. We also note the works \cite{TMF_24_1} -- \cite{TMF_24_2}, which showed the possibility of scalar halos in the external field of a scalar-charged black hole.

\TwoFigsReg{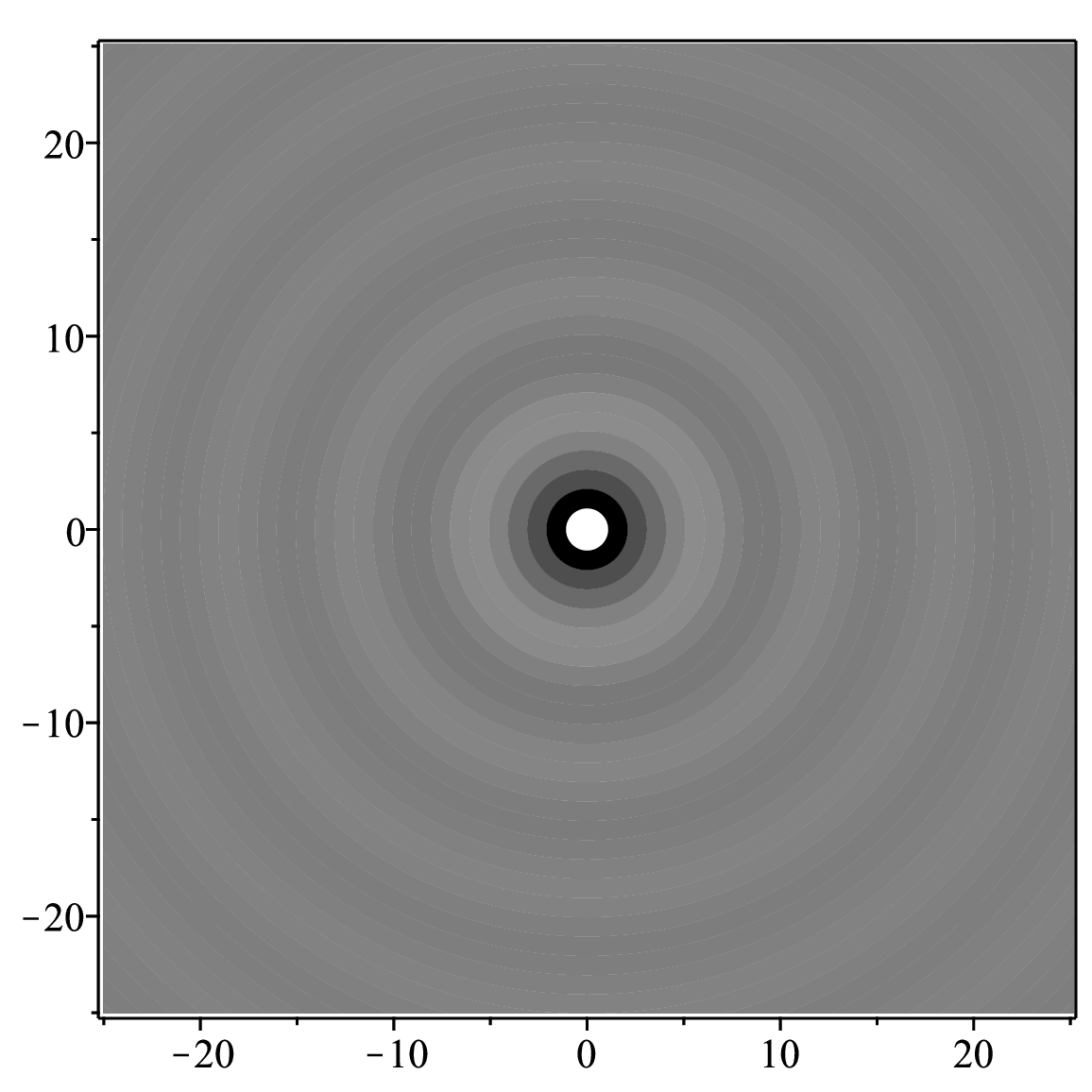}{7}{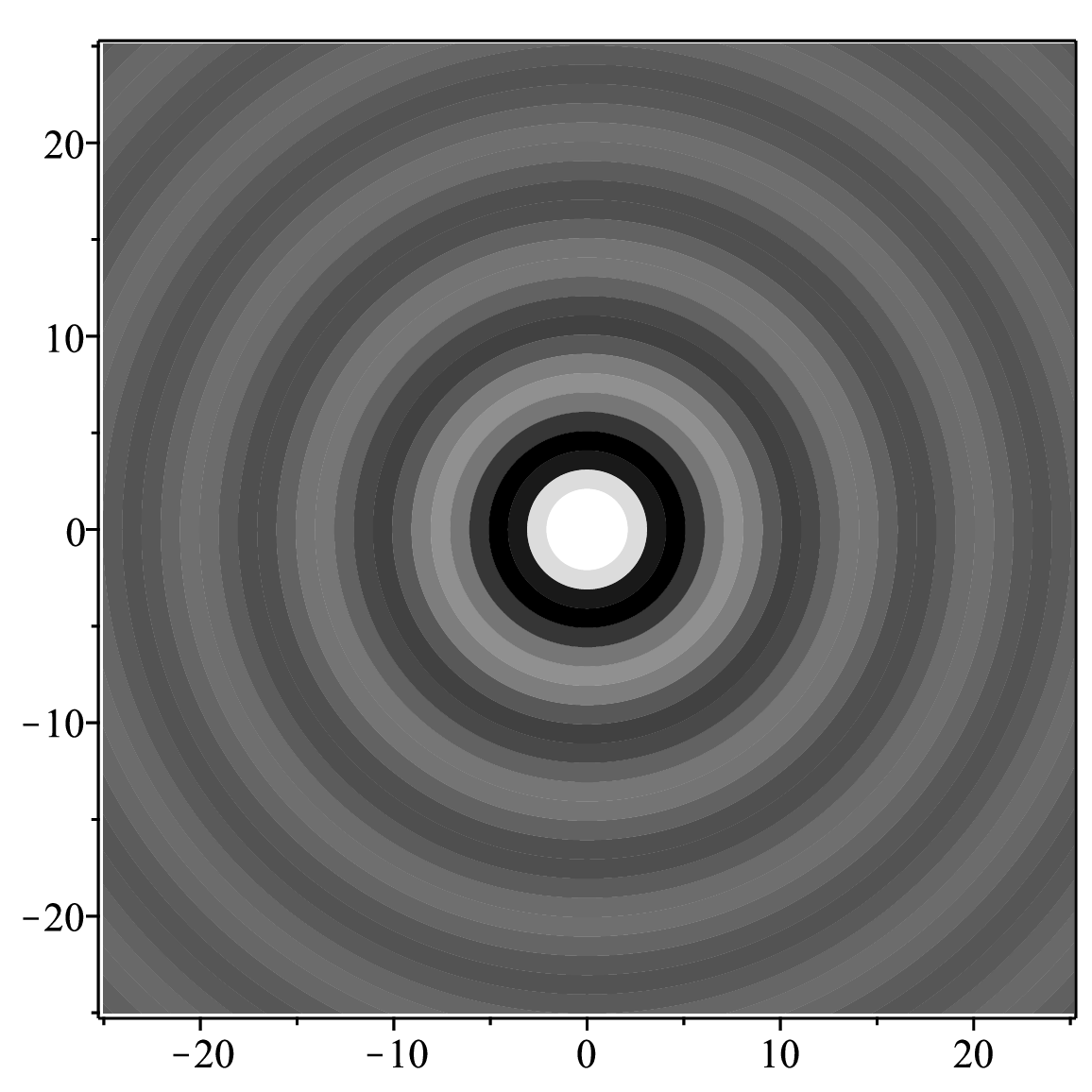}{7}{\label{fig9}Graph of $\Re(\phi_1)$ in the $P_0$ model, $t=0$}{\label{fig10}Graph of $\Re(\phi_1)$ in the $P_1$ model, $t=0$.}

\TwoFigsReg{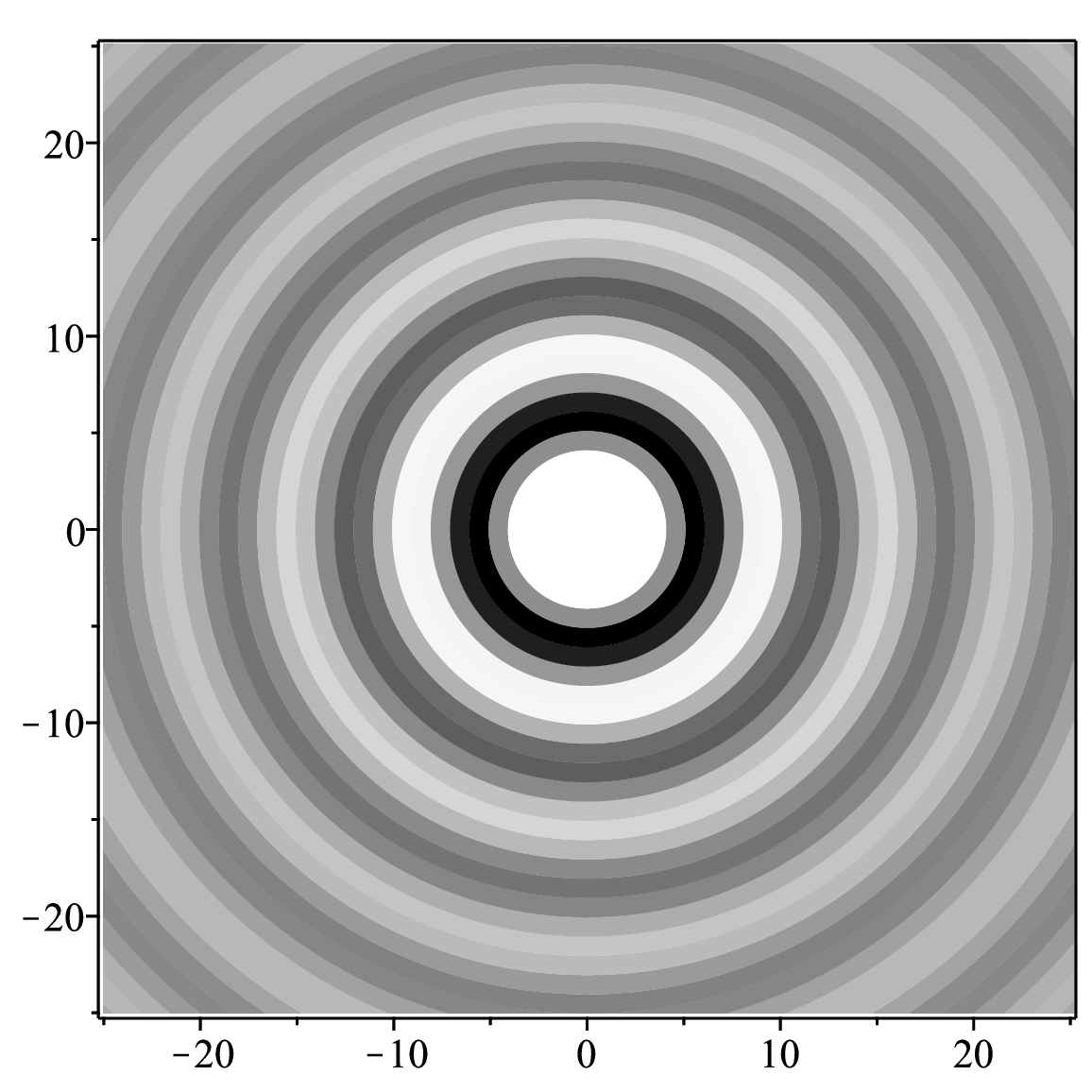}{7}{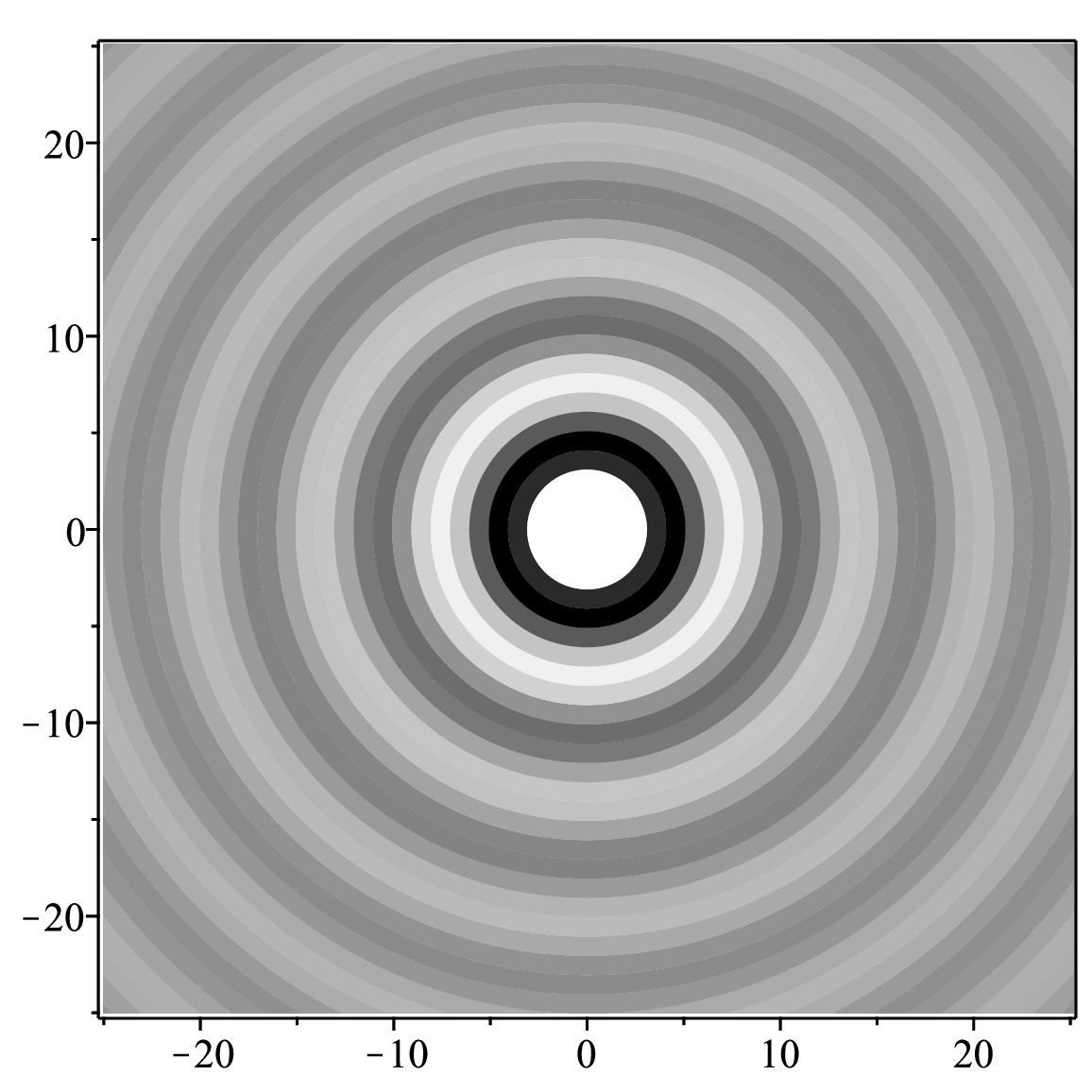}{7}{\label{fig11}Graph of $\Re(\phi_1)$ in the $P_0$ model, $t=30$}{\label{fig12}Graph of $\Re(\phi_1)$ in the $P_1$ model, $t=30$.}

In conclusion, we note that the above graphs demonstrate the general properties of the model under study for fundamental parameters of order unity in the Planck scales, while the real parameters, for example, in the field-theoretic SU(5) have much smaller values. To move to these parameters, it is necessary to apply a similarity transformation to the obtained results, using the invariant properties of the model under study \cite{Trans}.
Then we will obtain a distance between the halo strata of the order of $10^6$ in Planck units.

\section*{Acknowledgments}
Author is grateful to the participants of the seminar of the Department of Relativity and Gravitation Theory of Kazan University for useful discussion of the work. The author is especially grateful to professors S.V. Sushkov and A.B. Balakin for very useful comments and discussion of the features of modern modifications of the theory of gravity as applied to cosmology and astrophysics. The author is also grateful to the participants of the VI International Winter School-Seminar "Petrovskie Readings-2023" for a fruitful discussion of the report (11/29/23), which influenced the topic of the article, especially to Academician \fbox{A.A. Starobinsky} and Professor K.A. Bronnikov.

\section*{Founding}
The work was carried out at the expense of a subsidy allocated as part of the state support of Kazan (Volga Region) Federal University in order to increase its competitiveness among the world's leading scientific and educational centers.

\end{document}